\documentclass[12pt]{article}
\usepackage{graphicx}

\newcommand\pubnumber{}
\newcommand\pubdate{\today}

\def\okayama{Department of Physics\\
Okayama university, 3-1-1 Tsushimanaka, Kita-ku, Okayama, Japan}

\def\Title#1{\begin{center} {\Large #1 } \end{center}}
\def\Author#1{\begin{center}{ \sc #1} \end{center}}
\def\Address#1{\begin{center}{ \it #1} \end{center}}

\newcommand\pubblock{\rightline{\begin{tabular}{l} \pubnumber\\
         \pubdate  \end{tabular}}}
\newenvironment{Abstract}{\begin{quotation}  }{\end{quotation}}
\newenvironment{Presented}{\begin{quotation} \begin{center} 
             PRESENTED AT\end{center}\bigskip 
      \begin{center}\begin{large}}{\end{large}\end{center} \end{quotation}}
\def\Acknowledgements{\bigskip  \bigskip \begin{center} \begin{large}
             \bf ACKNOWLEDGEMENTS \end{large}\end{center}}

\def\beq{\begin{equation}}
\def\eeq#1{\label{#1}\end{equation}}
\def\eeqn{\end{equation}}

\def\beqa{\begin{eqnarray}}
\def\eeqa#1{\label{#1}\end{eqnarray}}
\def\eeqan{\end{eqnarray}}

\let\bar=\overbar

\def\Dslash{\not{\hbox{\kern-4pt $D$}}}
\def\dslash{\not{\hbox{\kern-2pt $\del$}}}

\def\msb{{\bar{\ssstyle M \kern -1pt S}}}

\begin{document}
\begin{titlepage}
\pubblock

\vfill
\Title{SOIKID, SOI pixel detector combined with superconducting detector KID}
\vfill
\Author{ Hirokazu Ishino, Atsuko Kibayashi, Yosuke Kida and Yousuke Yamada}
\Address{\okayama}
\vfill
\begin{Abstract}
We present the development status of the SOIKID, a detector combining
the SOI pixel detector and the superconducting detector KID (Kinetic Inductance
Detector).
The aim of the SOIKID is to measure X-ray photon energy with the
resolution better than that of the semiconductor detector.
The silicon substrate is used as the X-ray photon absorber.
The recoiled electron creates athermal phonons as well as the ionizing electron-hole 
pairs.
The KID formed at one side of the substrate surface detects the phonons
to measure the total energy deposited, 
while the SOI pixel detector formed on the other side of the substrate detects
the ionized carries to measure the position.
Combining the position and energy measurements, it is in principle possible to
have the extremely high energy resolution.
\end{Abstract}
\vfill
\begin{Presented}
SOIPIX 2015\\
Sendai, Japan,  June 3--4, 2015.
\end{Presented}
\vfill
\end{titlepage}
\def\thefootnote{\fnsymbol{footnote}}
\setcounter{footnote}{0}

\section{Introduction}

Semiconductor detectors have been widely used for measurements of X-ray 
photon energy in many scientific and industrial applications, as they
have a high energy resolution.
For some applications the resolution is not sufficient, however.
For example, the experiment operating at SPring-8 measures the electron
momentum in the material bulk to reveal the properties of the chemical bond
using the Compton-scattered photons~\cite{spring8}.
The energy resolution required for the experiment is 100~eV for 100~keV
photons, which is five times smaller than 500~eV of the semiconductor detectors.
Hence the experiment uses the X-ray dispersion crystal and a CCD camera
to measure the photon energy with the detection efficiency of 0.1\%.
Recently, Transition Edge Sensor (TES) micro-calorimeters have been developed
to measure the X-ray energy with the resolution of less than 50~eV for
100~keV photons~\cite{tes}.

On the other hand, M.~Kurakado {\it et al.} used Superconducting
Tunnel Junction (STJ) to detect phonons created in the substrate due to X-ray
absorption~\cite{kurakado}.
The phonons hitting at the substrate surface, where STJ is formed,
break Cooper pairs and generate quasi-particles.
The STJ has time response much faster than that of TES,
since the phonons that have energies greater than
the Cooper-pair binding energy can decay quickly.
The quasi-particles tunnel the SIS junction and are detected as the electric current.
M.~Kurakado {\it et al.} fabricated a series STJ detector and successfully 
detected 5.9~keV X-rays from a $^{55}$Fe source with the energy 
resolution of 70~eV, which is about half of that of the semiconductor
detector~\cite{kurakado}.

Kinetic Inductance Detector (KID)~\cite{kid} is one of the other options.
KID can be formed on a substrate with a single metal layer, reducing the
fabrication time and complexity compared with STJ.
In addition, KID can be readout with the frequency-domain, enabling us to have
the multiplexing readout of hundred channels with a single wire.
D.~C.~Moore {\it et al.} developed the KID made of aluminum thin film on
a silicon substrate. They obtained the energy resolution of 0.55~keV at
30~keV X-rays from a $^{129}$I source~\cite{jpl}. 
The advantage of using KID instead of STJ is the extendability;
the frequency domain multiplexing and the simple structure enable us to
extend the detector with a scale of $\sim$~kg target to search for
dark matter and neutrinoless double beta decays. 

\begin{figure}[htb]
\centering
\includegraphics[height=3in]{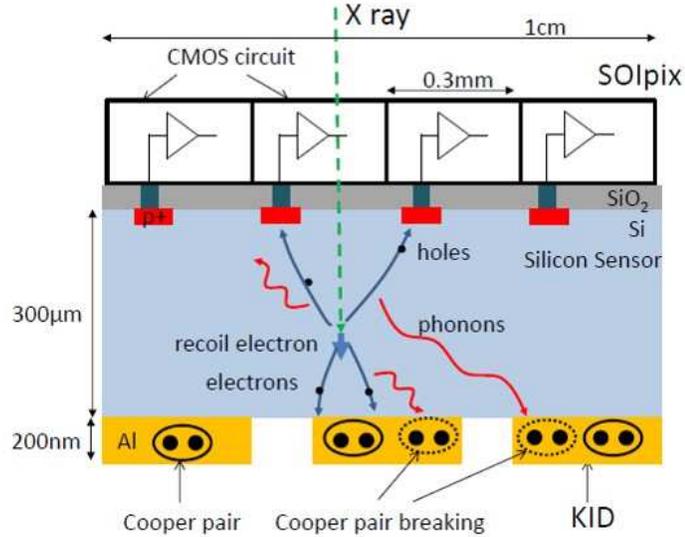}
\caption{Concept of SOIKID. The SOI pixel sensor detects the ionization signal
and measures the event position. The KID formed on the backside of the substrate
detects athermal phonons and measures the deposited energy.}
\label{fig:soikid}
\end{figure}

The energy resolution obtained by D.~C.~Moore {\it et al.} is measured
after the position corrections with the spatial resolution of 0.8~mm.
The position was identified using pulse height weighed mean.
If we could determine the event position within 0.1~mm, 
we would not only have a stringent corrections to the energy but also
can reject events occurring at the edge of the detector and 
expand the fiducial volume.
We propose SOIKID that combines a SOI (Silicon On Insulator)
pixel detector and a KID.
The concept of the detector is shown in Fig.~\ref{fig:soikid}.
The SOI pixel detector senses the ionization while the KID detects phonons.
A simultaneous detection of both ionization and phonons can provide 
a discrimination of the recoiled particle type, i.e., a silicon atom or an electron.
The backside of the SOI pixel is usually used as a bias pad formed
as a thin aluminum metal with a thickness of about 200~nm.
We propose to form a KID on this side.

In this proceedings, we give a status of the SOIKID development.
Details of the SOI pixel detector are given elsewhere~\cite{soi}.
We present the current status of the KID development, including
the detector performance studies and a readout system.
We note it has been found that the SOI pixel can function at the
low temperature of less than 1~K at which the superconducting detector
can work. This is the motivation of using SOI pixel detector as the
ionization detection and the position measurements.
The behavior of the MOSFET at the low temperature is different from
that at the room temperature.
The characterization of the behavior is being studied.

\section{Development of KID for phonon detections}

\begin{figure}[htb]
\centering
\includegraphics[height=1.5in]{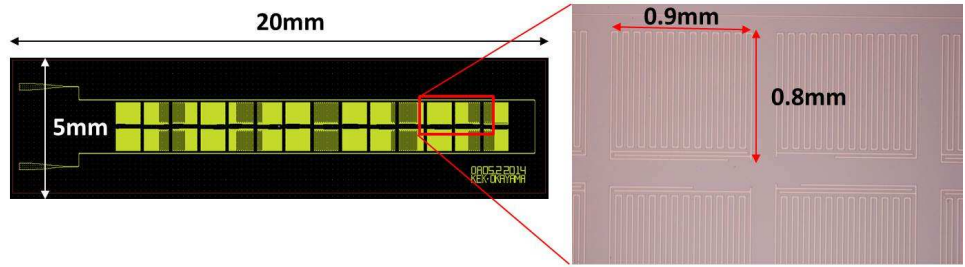}
\caption{(Left)The KID detector design.
Twenty-six resonators are formed with the feed-line on 
$5\times 20$~mm$^2$ silicon substrate.
(Right)Picture of the fabricated KID. The size of one resonator
is also shown.
}
\label{fig:kid}
\end{figure}

We designed an array of a KID as shown in Fig.~\ref{fig:kid}.
The detector consists of 26 resonators.
One resonator is composed of an inductance formed as a meander 
structure and an interdigital capacitor.
The two components form a $LC$ resonant circuit tuned to have
a resonant frequency in the microwave range of $4\sim6$~GHz.
The frequency range matches the bandwidth of a cryogenic HEMT
amplifier.
The 26 resonators have different resonant frequencies with a frequency
spacing of 15~MHz.
The resonators are inductively coupled to the feed-line through which
we send microwaves having the resonant frequencies.
In a stable condition, we expect no output of the injected microwaves, as
those microwaves are absorbed by the resonators.
When one of the resonators has some energy deposited, Cooper-pairs are broken
on the resonator.
Then, the kinetic inductance changes, where the kinetic inductance is 
the inductance induced by the kinetic energy of Cooper pairs.
Therefore the resonator's resonance frequency changes, resulting in the
output signal of the microwave.
By measuring the amplitude and phase of the microwave output,
we can measure the deposited energy.

The design of the KID is tuned by carefully taking into account the cross-talk;
the tuned parameters include distances between resonators and the feed-line,
and
the deployment of resonators such that those with close resonant frequencies
are not in the vicinities.
When we separate two resonators at a larger distance, cross-talks 
between them become
small, but the detection efficiency of athermal phonons decreases as the
surface coverage area diminishes.
We iterated the design using an electromagnetic simulation (SONNET),
fabrication of a detector using a Nb film and evaluations.
The Nb detector can be evaluated quickly as Nb becomes superconductor below
a temperature of 9.2~K, enabling us to use liquid helium with pumping.
The distance between the feed-line and the resonators also determine the
coupling quality factor.
We tuned the distance so that the total quality factor has the order of $10^4$.

For the detection of athermal phonons, materials that have the lower critical 
temperature are better, since the Cooper-pair binding energy is proportional to
the critical temperature according to the BCS theory,
$2\Delta = 3.52k_{\rm B}T_{\rm c}$, where
$2\Delta$ is the Cooper-pair binding energy corresponding to the gap energy
in the superconductor, $k_{\rm B}$ is the Boltzmann constant and $T_{\rm c}$
is the critical temperature~\cite{bcs}.
Aluminum has the critical temperature of 1.2~K and has $2\Delta=0.34$~meV.
The lower the critical temperature is, the more the KID has a sensitivity to the
phonons, as the phonon population is larger for lower phonon energies.
However, it may be difficult to operate KID made of the material having the lower critical temperature, since we need a powerful refrigerator. 
The candidate materials are Al, Ti, TiN~\cite{jpl}, that can be operated 
at $0.1\sim 0.3$~K.
We choose Al as the KID material, as the SOI pixel detector has Al thin layer
on the backside of the substrate, and also an Al thin layer can be formed using
a sputter machine in the KEK clean room.

\begin{figure}[htb]
\centering
\includegraphics[height=3in]{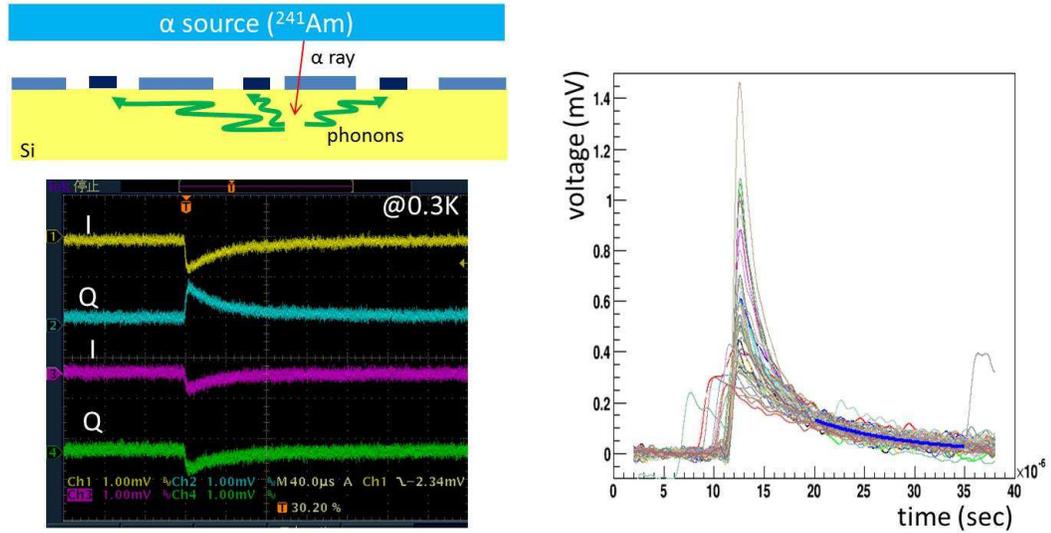}
\caption{(Left upper)Experimental setup to irradiate $\alpha$ particles
from a sealed $^{241}$Am source. (Left lower) Oscilloscope image
of phonon signals for two resonators. (Right) Superimposition of 50 phonon
signals of a single resonator.}
\label{fig:phonon-signal}
\end{figure}

We fabricated KID with a 200~nm Al layer on $5\times 20$~mm$^2$
high purity silicon substrate with a thickness of 300~$\mu$m.
The device was mounted on a ceramic plate by applying the GE7031 varnish,
and covered with a brass housing.
In order to detect athermal phonons, we placed a sealed $^{241}$Am source
on the housing to irradiate $\alpha$ particles with the energy of 5.5~MeV.
The detector and the source were placed inside the He-3 sorption refrigerator 
and cooled down to 0.3~K.
Figure~\ref{fig:phonon-signal} shows an experimental set up and an oscilloscope
image of the phonon signals for two resonators.
The coincidence of the signal is result of the athermal phonon signals from
the substrate.
We also show superimpositions of 50 phonon signals for a single resonator.
The pulse height corresponds to the location of the $\alpha$ particle 
irradiation position, i.e., the larger the signal height, the closer the position.
Although the initial part of the signal is different, the signal height becomes
almost the same for the time $>20$~$\mu$sec.
This is supposed to be the thermalization of the substrate; 
the athermal phonons propagate to all the region of the substrate
and the athermal phonon density became almost uniform.
The same phenomenon was seen in ~\cite{jpl}.
The decay time attributes to either the athermal phonons decay or the
phonons escape to the ceramic holder.
The decay time is measured to be $9.3\pm0.4~\mu$s, comparable to
the estimated decay time obtained from the thermal capacitance of the 
substrate and the thermal conductance of the varnish.

\begin{figure}[htb]
\centering
\includegraphics[height=1.6in]{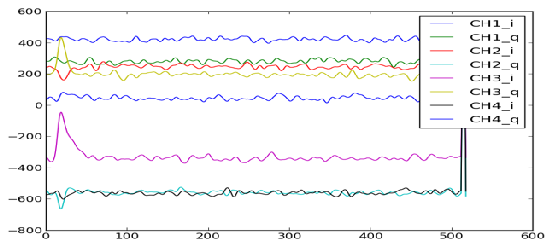}
\includegraphics[height=1.5in]{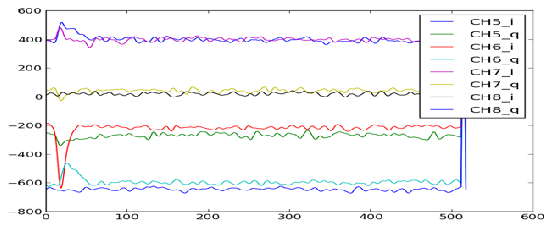}
\caption{
KID pulse signals for 8 resonators readout using
the Kintex-7 evaluation board with the customized firmware.
}
\label{fig:kid-signal}
\end{figure}

We have developed a customized KID readout system.
We employ the KC705 Kintex-7 evaluation board
for the firmware development.
The board has a card consisting of both ADC and DAC with which we can generate
a resonant frequency comb and measure the KID output to monitor
changes of the amplitude and the phase.
In the firmware, the input signal is demodulated and converted
to the time-domain signal for each resonator.
The time ordered signal is divided into two; one is sent to a trigger module
where the time ordered signals from all the channels are added to generate
a trigger signal in case the total signal exceeds a threshold, 
and the other is sent to a FIFO where, in case trigger is issued, the
data are sent to the buffer or deleted otherwise.
The data sent to the buffer are transfered to a PC through the SiTCP~\cite{sitcp}.
Figure~\ref{fig:kid-signal} demonstrates the KID pulse signals for 8 channels
obtained using the readout system.

\section{Specification of SOI pixel detector}

The SOI pixel detector is required to have position resolution of 0.1~mm
and to work at a temperature of 0.3~K.
The former requirement is achieved with 0.3~mm pixel size.
We plan to have $30\times 30$ pixels on the $10\times 10$~mm$^2$ substrate,
and readout the pixels for $x$ and $y$ strips serially to allow a small
number of cables inside the refrigerator that can reduce the thermal injection
from the room temperature.
The pixel signal is digital.
The readout period is 1~kHz at maximum, so one strip line readout takes
16~$\mu$sec.
The latter requires us to operate the detector with lower power consumption,
0.1~$\mu$W per pixel, which is challenging.
The signal to noise ratio is required to be greater than 4 for 6~keV energy deposition
in the substrate.
The requirement comes from the rejection of accidental noise hit 
at one per thousand pixels.
The noise magnitude is required to be less than 100 electrons.

\section{Summary}

We propose SOIKID that is a combination of  a SOI pixel detector
and a KID.
The former determines the event position by sensing the ionization signal,
while the latter measures the deposit energy by detecting athermal phonons.
We have demonstrated the KID made of Al could detects athermal phonon.
The design of the KID has been tuned to have small cross-talks and 
a large covering area to have a higher detection efficiency of phonons.
The inductive coupling to the feed-line has been optimized 
to have a quality factor of the order of $10^4$.
A customized KID readout system is being developed using a 
Kintex-7 evaluation board.
We have successfully readout pulse signals from 8 resonators 
simultaneously.
We have identified the specification of the SOI pixel detector that operates
at a temperature of 0.3~K in the refrigerator.



\Acknowledgements
The authors thank Prof. Y.~Arai at KEK and  Prof. S.~Kawahito at Shizuoka university.
The authors would like to thank KEK Detector Technology Project
for the support of the development of superconducting detectors at KEK.
This work is supported by the Grant-in-Aid for Scientific Research
on Innovative Area No. 26109508.

\end{document}